\documentclass[12pt, a4paper]{article}
\pdfoutput=1
\usepackage[dvipdfmx]{graphicx}
\usepackage{amssymb}
\usepackage{amsmath}
\usepackage{bm}
\usepackage{color}
\usepackage{cite}
\usepackage{epstopdf}            
\usepackage{epsfig}
            
\newcommand{\beq}{\begin{equation}}
\newcommand{\eeq}{\end{equation}}

\usepackage[colorlinks=true, linkcolor=black, citecolor=black,
urlcolor=black]{hyperref}

\begin{document}

\begin{titlepage}

\begin{flushright}
\begin{minipage}{0.2\linewidth}
\normalsize
WU-HEP-16-22 \\*[50pt]
\end{minipage}
\end{flushright}

\begin{center}

{\Large
{\bf 
Dark matter physics, flavor physics  \\
and LHC constraints \\
 in the dark matter model with a bottom partner
}
}

\vskip 2cm

Tomohiro Abe$^{1,2}$,
Junichiro Kawamura$^{3}$, 
Shohei Okawa$^4$
and
Yuji Omura$^{2}$

\vskip 0.5cm

{\it $^1$
Institute for Advanced Research, Nagoya University, Nagoya 464-8602, Japan}\\[3pt]
{\it $^2$
Kobayashi-Maskawa Institute for the Origin of Particles and the
Universe, \\ Nagoya University, Nagoya 464-8602, Japan}\\[3pt]
{\it $^3$Department of Physics, Waseda University, 
Tokyo 169-8555, Japan
}\\[3pt]
{\it $^4$
Department of Physics, Nagoya University, Nagoya 464-8602, Japan}\\[3pt]

\vskip 1.5cm

\begin{abstract}
In the scenario that dark matter (DM) is a weakly interacting massive particle,
there are many possibilities of the interactions with the Standard Model (SM) particles to achieve
the relic density of DM. 
In this paper, we consider a simple DM model where the DM candidate is a complex scalar boson.
The model contains a new complex gauge singlet scalar boson and a new fermion whose gauge charge is the same
as the right-handed down-type quark. We dub the new fermion the bottom partner. 
These new particles have Yukawa interactions with the SM down-type quarks.
The DM candidate interacts with the SM particles through the Yukawa interactions.
The Yukawa interactions are not only relevant to the annihilation process of the DM
but also contribute to the flavor physics, such as the $\Delta F=2$ processes. 
In addition, the flavor alignment of the Yukawa couplings is related to the decay modes of the bottom partner,
and thus we can find the explicit correlations among the physical observables in DM physics, flavor physics, and the signals at the LHC. 
We survey the $\Delta F=2$ processes 
based on the numerical analyses of the thermal relic density, the direct detection of the DM, and the current LHC bounds.
We investigate the perturbative bound on the Yukawa coupling as well. 
A Study of a fermionic DM model with extra scalar quarks is also given for comparison. 
\end{abstract}

\end{center}
\end{titlepage}

\section{Introduction}
\label{sec;intro}
The cosmological observations exhibit the existence of dark sector in our universe.
In particular, the existence of dark matter (DM) is supported by many independent observations.
The WMAP and the Planck experiments have shown that the relic density
of DM is about 5 times larger than the one of the baryon in our universe \cite{DMexperiment2,DMexperiment}.
Such a large DM density implies the existence of physics beyond the standard model (SM)
and is driving particle physicists to build the Beyond Standard Models (BSMs). 

A lot of ideas motivated by DM have been proposed so far.
One simple popular idea is as follows. DM candidates are neutral under the electromagnetic and SU(3)$_c$ symmetries,
and interact with the SM particles via the electroweak gauge couplings. 
In this case, the DM is charged under the SU(2)$_L \times$U(1)$_{Y}$ gauge symmetry 
and the mass of the DM is an unique parameter to explain the relic density.\footnote{
See, for instance, Ref. \cite{IDM}.}

There is another possibility that DM candidates are singlet under the SM gauge groups. The SM gauge symmetry itself would not prohibit couplings between DM and the SM Higgs boson. 
In addition, DM may have Yukawa couplings with the SM particles.
If DM is a complex scalar field,
such a Yukawa coupling is realized by introducing extra quarks or leptons.
This kind of model would be one of the best candidates for the BSM,
because many observables are not changed  drastically from the SM predictions in this model  \cite{vectorlike,vectorlike2,vectorlike3,vectorlike31,vectorlike4,Kawamura,vectorlike5}.
Besides, it is also interesting that such a simple model can be tested not only in dark matter physics and the LHC experiments but also in
flavor physics.\footnote{The correlations have been discussed in the models with flavor symmetry \cite{FlavoredDM1,FlavoredDM2,FlavoredDM3}.} 
The Yukawa couplings between the DM and the SM particles 
are generally flavor-dependent, so that the contributions to flavor changing processes 
modify the SM predictions. 
If DM is discovered in the LHC and DM experiments in future, 
we have to test a lot of DM models in many observables.
In this simple model, we could find some correlations among some observables and give some
explicit predictions to the results in future experiments.


Based on this consideration, in this paper, we concentrate on a DM model with a scalar DM candidate ($X$) and a vector-like quark ($F$) whose SM charges are the same as the ones of the right-handed down-type quarks. 
The vector-like quark can be easily introduced without any gauge anomaly.
$X$ is a complex scalar and neutral under the SM gauge groups.
It couples with right-handed down-type quarks ($d^i_{R}$) via the Yukawa couplings involving $F$:  $\lambda_{ i} \overline{F_L}  X^\dagger d^i_{R}$ ($i=d, \, s, \,b$). 
Besides, $X$ interacts with the SM Higgs doublet, denoted by $H$, via a 4-point coupling, $\lambda_H |X|^2|H|^2$.
Note that this setup differs from the models in Refs. \cite{vectorlike,vectorlike2,vectorlike3,vectorlike31,vectorlike4,vectorlike5} 
and we give the integrated quantitative study including flavor physics, which has not been done in Refs. \cite{vectorlike,vectorlike2,vectorlike3,vectorlike31,vectorlike4,Kawamura,vectorlike5}.

The recent indirect and direct detection experiments of DM succeed in drawing strong bounds on simplified DM models. 
For instance, the stringent bounds on the setup from the AMS-02 experiment \cite{AMS-02} are recently suggested in Refs. \cite{Cuoco:2016eej,Cui:2016ppb}. If the DM mass is below 1 TeV, the annihilation cross section associated with $b \overline{b}$ and $W^+W^-$ in the
final state should be below the required value for the thermal relic density.
Besides, the direct detections of DM at the LUX \cite{LUX2015,LUX2016} and Panda \cite{Panda} experiments strongly 
limits the interaction of DM with light quarks. Actually, if $\lambda_H$ is large and the DM mass is below 1 TeV, the recent results of the LUX experiment \cite{LUX2016} have already excluded our DM model.  
In our setup, however, the DM can evade the strong bounds
from the indirect and direct detections because of the suppressed $s$-wave contribution to the annihilation
and the alignment of the Yukawa couplings assuming small $\lambda_H$.
In this sense, our model is one of the realistic DM models which could be tested by many independent observables near future as mentioned below.  
Moreover, it is also interesting that this kind of model could be effectively realized in the framework of the grand unified theory \cite{Ko:2016lai}. This is also our motivation to consider this simplified DM model.

The outline of our analysis is as follows.
In Sec. \ref{sec;DM}, we will show that the $F$ exchanging diagrams dominantly
contribute to the DM annihilation process for the thermal relic density.
We need large Yukawa couplings, $\lambda_i$, because the $s$-wave contribution 
in the annihilation process is suppressed by the down-type quark masses. 
Besides, the relation, $\lambda_b \gg \lambda_{d,s}$, should be satisfied to evade
the strong bounds from the flavor physics and the LHC experiments as well as the direct detection of the DM.
Such a hierarchy of $\lambda_i$ leads that $F$ dominantly decays to the bottom quark and the DM, $X$. 
In this sense, we call $F$ a "$bottom$ $partner$."
Eventually, only three parameters are relevant to the DM physics and LHC experiments: namely,
masses of $F$ and $X$, and $\lambda_b$. 

In our model, we do not assign any flavor symmetry,
so that $\lambda_d$ and $\lambda_s$ cannot be vanishing, whereas $\lambda_b \gg \lambda_{d,s}$ is assumed.
In general, flavor physics is very sensitive to the contributions of new physics, even if the new physics scale is above TeV-scale.
Then the (future) flavor experiments are expected to be sensitive to our model, even if $\lambda_d$ and $\lambda_s$ are tiny.   
We discuss the $\Delta F=2$ processes based on our results in the dark matter and LHC physics, in Sec. \ref{sec;flavor}.  
Then, we find correlations among $\Delta F=2$ processes, DM observables and the LHC signals, as will be discussed in Section \ref{sec;flavor}.  
We suggest that our simple model can be tested by precise measurements of the $\Delta F=2$ processes, once the DM is discovered by the LHC experiments and/or DM observations.  
This is a main goal of this work. 
In Sec. \ref{sec;fermionicDM}, we also compare our results with the ones in another setup, 
where there are a fermionic DM ($ \widetilde X$) and a extra scalar quark ($\widetilde F$) instead of $X$ and $F$.
$ \widetilde X$ is a Dirac fermion in our study.

In Sec. \ref{sec;setup}, we introduce the setup of our model.
Then, we study the signals of the DM and the extra quark in the LHC experiments, dark matter physics and flavor physics, in Secs. \ref{sec;LHC}, \ref{sec;DM} and \ref{sec;flavor} respectively. The triviality bound on the Yukawa couplings
is studied, in Sec. \ref{sec;Triviality bound}.
In Sec. \ref{sec;fermionicDM}, we discuss 
another setup that a fermionic DM ($ \widetilde X$) and a extra scalar quark ($\widetilde F$) are introduced
instead of $X$ and $F$, and compare our predictions in both cases.  
Section \ref{sec;conclusion} is devoted to the summary.

\section{Setup}
\label{sec;setup}
In this section, we introduce our model with a vector-like quark. 
Similar setup has been proposed, motivated by the DM physics and the LHC physics \cite{vectorlike,vectorlike2,vectorlike3,vectorlike31,vectorlike4, Kawamura, vectorlike5}. 

We introduce an extra down-type quark, $F$, 
carrying SM charges as in Table \ref{table1}.
\begin{table}[htb]
 \label{table1}
\begin{center}
\begin{tabular}{cccccc}
\hline
\hline
Fields & ~~spin~~   & ~~$\text{SU}(3)_c$~~ & ~~$\text{SU}(2)_L$~~  & ~~$\text{U}(1)_Y$~~ &~~U(1)$_{X}$~~     \\ \hline  
   $F$  & 1/2 & ${\bf 3}$        &${\bf1}$      &         $-1/3$      & $1$          \\
   $ X$  & 0 & ${\bf1}$         &${\bf 1}$      &            $0$   & $-1$         \\ \hline \hline
\end{tabular}
\end{center}
\caption{Extra fields in our model with global U(1)$_{X}$. }
\end{table} 
$F$ is a Dirac fermion and the charge assignment is the same as the one of the right-handed down-type quarks, $d^i_R$ ($i=1,\,2,\,3$). In our notation, $(d^1, \, d^2, \, d^3)$ correspond to $(d, \, s, \, b)$.
We assign a global U(1)$_{X}$ charge to $F$ to distinguish it from the SM down-type quarks. 
In addition, we introduce a complex scalar, denoted by $X$, which is also charged under the U(1)$_{X}$ symmetry.
$X$ is stable thanks to the U(1)$_X$ symmetry and is a DM candidate in our model.
The charge assignment is summarized in Table \ref{table1}.

Now we can write down the potential for the extra quark and the scalar:
\begin{eqnarray}
V&=&V_{\rm F}+V_{\rm X}, \\
V_{\rm F}&=& m_F \overline{F_L}F_R + \lambda_{ i} \overline{F_L}  X^\dagger d^i_{R} + h.c., \label{eq;VB}  \\
V_{\rm X}&=& m^2_X |X|^2+ \lambda_H |X|^2 |H|^2 + \lambda_X |X|^4 -m^2_H |H|^2+ \lambda |H|^4. \label{eq;VX} 
\end{eqnarray}
Each of the Yukawa couplings, $ \lambda_{ i}$, induces the decay of $F$: $F \to X^\dagger \, d^i$.
Note that $V_{\rm X}$ includes the coupling of $X$ to the SM Higgs boson ($H$).
$\lambda_H$ plays a crucial role in dark matter physics, as discussed in Sec. \ref{sec;DM}.

\section{Phenomenology }
\label{sec;phenomenology}
In our model, there are several parameters that can be determined 
by combining the analyses of dark matter physics, flavor physics and
the direct searches at the LHC. The relevant parameters in our study are as follows:
\beq
\label{parameters}
m_X, ~m_F,~\lambda_H,~\lambda_b,~ Re(\lambda_s),~Im(\lambda_s),~ Re(\lambda_d),~Im(\lambda_d).
\eeq
Note that we can define $\lambda_b$ as a real one, without loss of generality. 
In order to avoid the stringent bounds from flavor physics and direct detections of the DM,
we assume the following relation,
\beq
|\lambda_b| \gg |\lambda_d|, \, |\lambda_s|.
\eeq
In this case, $F$ mainly decays to $X$ and the bottom quark, 
and the dominant annihilation process of $X$ is $X \, X^\dagger \to \overline{b} \, b$ through the $t$-channel exchange of $F$, 
as far as $\lambda_H$ is relatively small.
This means that $\lambda_b$, as well as $m_X$ and $m_F$, can be fixed by the direct search for $F$ and $X$
in the $b \overline{b}$ signal accompanied by the large missing energy at the LHC and the DM observables, i.e., 
the relic abundance and the direct/indirect detections.

On the other hand, $\lambda_d$ and $\lambda_s$ are tiny in our setup, but not vanishing in general.
The Yukawa couplings are strongly constrained by flavor physics and should be less than ${\cal O}(0.01)$,
as we see in Sec. \ref{sec;flavor}. In other words, we can expect the sizable deviations in 
physical observables in flavor violating processes. In fact, we will find some correlations 
among the observables in the $\Delta F=2$ processes and derive explicit predictions for them, taking the analyses of the DM and LHC physics into account, in Sec. \ref{sec;flavor}.

\subsection{Constraints from the direct searches at the LHC }
\label{sec;LHC}

First, let us discuss the collider bounds from the new physics searches.
In our model, the extra quark ($F$), which we call a $bottom$ $partner$, is produced at the LHC and mainly decays into a bottom quark and  DM, via the Yukawa coupling, $\lambda_b$. 
This signal is similar to the one in supersymmetric models: that is,  $bb+E_T^{\rm miss}$.
In order to extract the exclusion limit for the bottom partner, 
we generate the UFO model file~\cite{Degrande:2011ua} using FeynRules~\cite{Alloul:2013bka}.
We use the MadGraph5~\cite{Alwall:2014hca} to simulate signal events with a pair produced vector-like quarks at the leading order (LO) with up to a parton.  
The generated events are passed into PYTHIA6~\cite{Sjostrand:2006za} and DELPHES3~\cite{deFavereau:2013fsa} to accommodate parton showering and fast detector simulation. 
The matrix element is matched to parton showers according to the MLM scheme~\cite{Caravaglios:1998yr}. The generated hadrons are clustered using the anti-$k_T$ algorithm~\cite{Cacciari:2008gp} with the radius parameter $\Delta R = 0.4$. 
In the analysis for the $bb + E_T^{\rm miss}$ search, 
we assume that the b-tagging efficiency obeys
a formula $0.80\times\tanh(0.003p_T)\times 30/(1+0.086 p_T)$
which is employed in the ATLAS DELPHES card in the MadGraph5,  
then we rescale the event weight by multiplying a factor of 1.2 
to emulate the experimental b-tagging efficiency where the working point is $77\%$
for $t\bar{t}$ events.

\begin{figure}[!t]
\centering
{\epsfig{figure=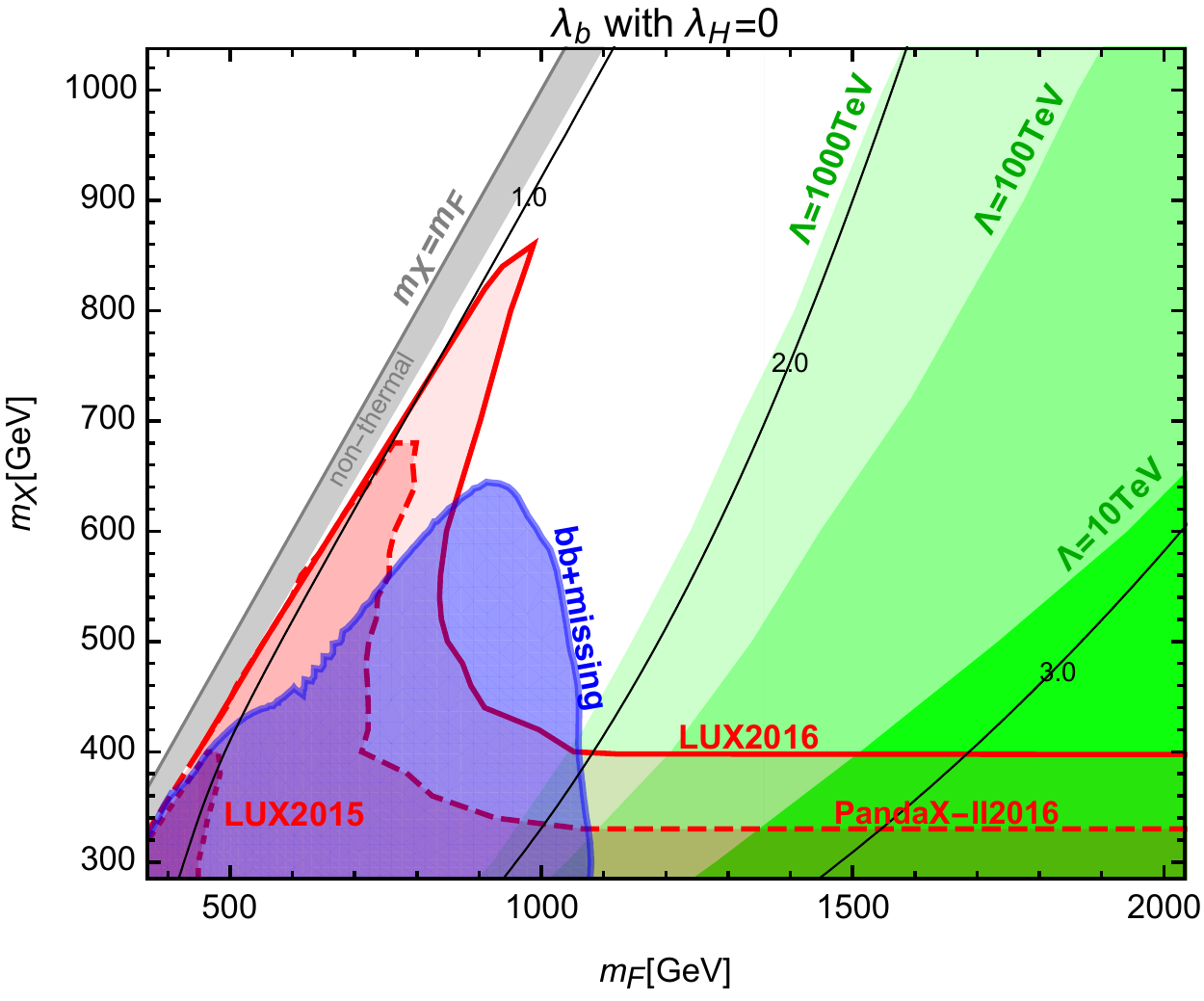,width=0.5\textwidth}}{\epsfig{figure=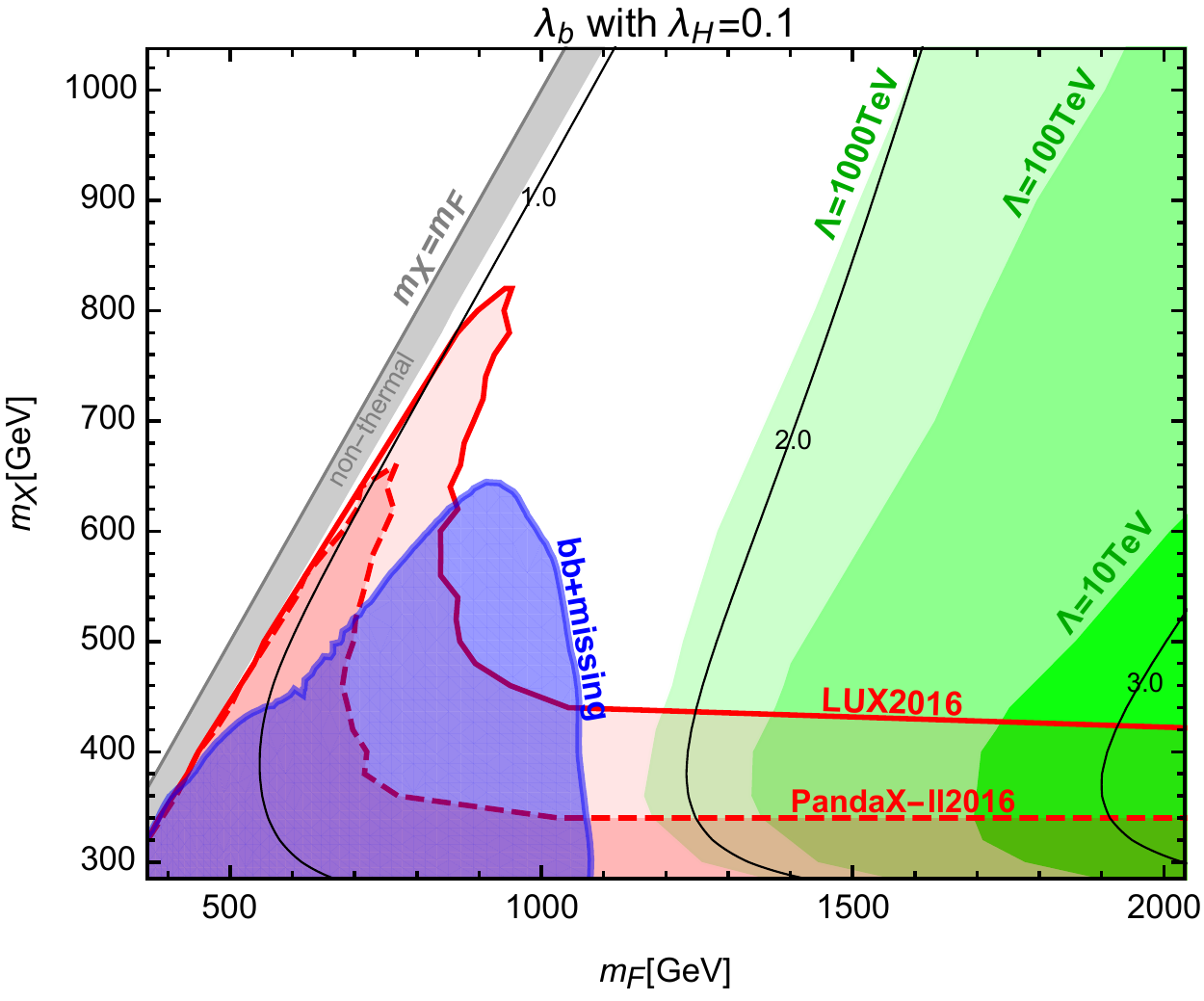,width=0.5\textwidth}}
\vspace{-0.5cm}
\caption{The required values of $\lambda_b$ for $\Omega h^2=0.1198 \pm 0.0015$. The blue regions are excluded by the
$bb+E_T^{\rm miss}$ at the LHC. The red regions are excluded by the direct detection experiments. The Yukawa couplings diverge below 1000 TeV in the green regions. 
 }
\label{fig1}
\end{figure}

Following the analysis of $bb + E_T^{\rm miss}$ in Ref.~\cite{bbMET}, we draw the exclusion limit in Fig. \ref{fig1}.
This result is given referring the latest data of the LHC Run-II with $\sqrt{s} = 13$ TeV and 3.2 $\rm fb^{-1}$.
We compared the expected number of events in each signal region
defined in the ATLAS analysis
with their 95$\%$ C.L exclusion limits shown in the Ref.~\cite{bbMET}.   
The Yukawa couplings $\lambda_i$ potentially
induce large production cross section of the extra-quark pairs 
by the t-channel process mediated by $X$, 
but this process is suppressed
by the parton distribution function in our case with 
$\lambda_b \gg \lambda_{s,d}$.

\subsection{Dark Matter Physics }
\label{sec;DM}
Next, we survey the relic abundance and the direct/indirect detection of $X$.

\subsubsection{Relic DM abundance}

Annihilation and coannihilation of $X$ and $X^\dagger$ to the SM particles should be sufficiently large
in order to be consistent with the cosmological observation of the DM abundance in our universe.
Those processes are governed by the $t$-channel exchange of $F$ and
the $s$-channel exchange of the Higgs. 
The $t$-channel process is dominant for small $\lambda_H$.
We employ {\tt micrOMEGAs\_3.6.9.2} \cite{micromegas} to evaluate the relic abundance of the DM. 
In our numerical analysis, we use the value reported by the Planck collaboration: $\Omega h^2=0.1198 \pm 0.0015$ \cite{DMexperiment}.

\begin{figure}[!h]
\centering
{\epsfig{figure=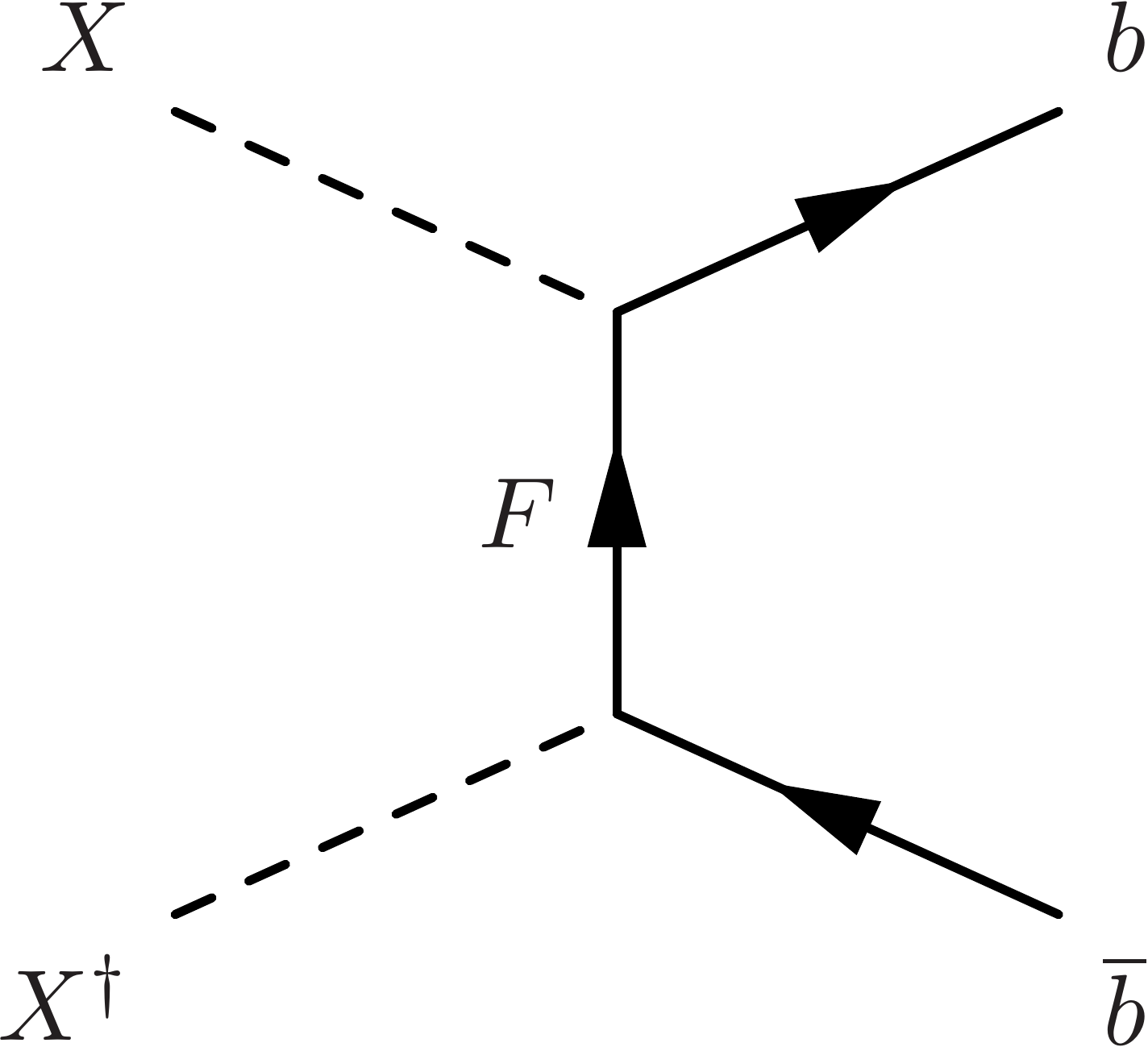,width=0.3\textwidth}}
\caption{The dominant process of DM annihilation in the $t$-channel.
 }
\label{diagram1}
\end{figure}

The main process of the DM annihilation is $X \, X^\dagger  \to  b \, \overline b$ in the $t$-channel shown in Fig. \ref{diagram1}. The Yukawa coupling, $\lambda_{ b} \overline{F_L}  X^\dagger b_{R}$,
depends on the chiralities of  $F$ and $b$, so that the bottom quark in the final state is right-handed in the massless limit.
Then, we conclude that the $s$-wave contribution of the annihilation is strongly suppressed
by the bottom quark mass, so that $\lambda_b$ is required to be ${\cal O}$(1) as far as $\lambda_H$ is small
and coannihilation is inefficient.

The coannihilation processes, such as $F \, \overline{F}  \to  g \,g$, 
may drastically decrease the relic density if $m_F$ is close to $m_X$. 
In fact, the coannihilation contribution leads too small relic density of $X$ 
in the region with $m_F \approx m_X$. The gray region in Fig. \ref{fig1} corresponds to the one that predicts $\Omega h^2 < 0.1183$ that is out of the $1 \sigma$ region of the Planck result. Our analysis for the DM density includes $X \, X^\dagger  \to  b \, \overline b \, g$ in addition to the annihilation and the coannihilation. 
This process has a non-negligible contribution for  $r \equiv m_F / m_X \lesssim 2$ and 
dominates over the $t$-channel process for $r \lesssim 1.12$. Our analysis only includes the $s$-wave contribution of this process. 

We show the required values of $\lambda_b$ in Fig. \ref{fig1}, where $\lambda_H=0$ (left panel) and $\lambda_H=0.1$ (right panel), respectively.
On the solid lines, $\lambda_b$ is fixed at $1, \,2, \,3$ from left to right respectively, and the relic density satisfies the observed value. 
In the compressed region below $m_F=1$ TeV, the mass difference between $m_X$ and $m_F$ is about $50$ GeV
to satisfy the Planck data within $1 \sigma$. As mentioned above, we see that $\lambda_b$ is required to be ${\cal O}$(1) 
in the most of parameter region.

\subsubsection{Direct/indirect detections of DM}
\label{sec;DMdirect}

DM can be detected by the observation of the DM scattering with nuclei.
Recently, we can successfully draw the stringent exclusion lines thanks to a lot of efforts of
the LUX \cite{LUX2015,LUX2016} and Panda \cite{Panda} collaborations.

\begin{figure}[!h]
\centering
{\epsfig{figure=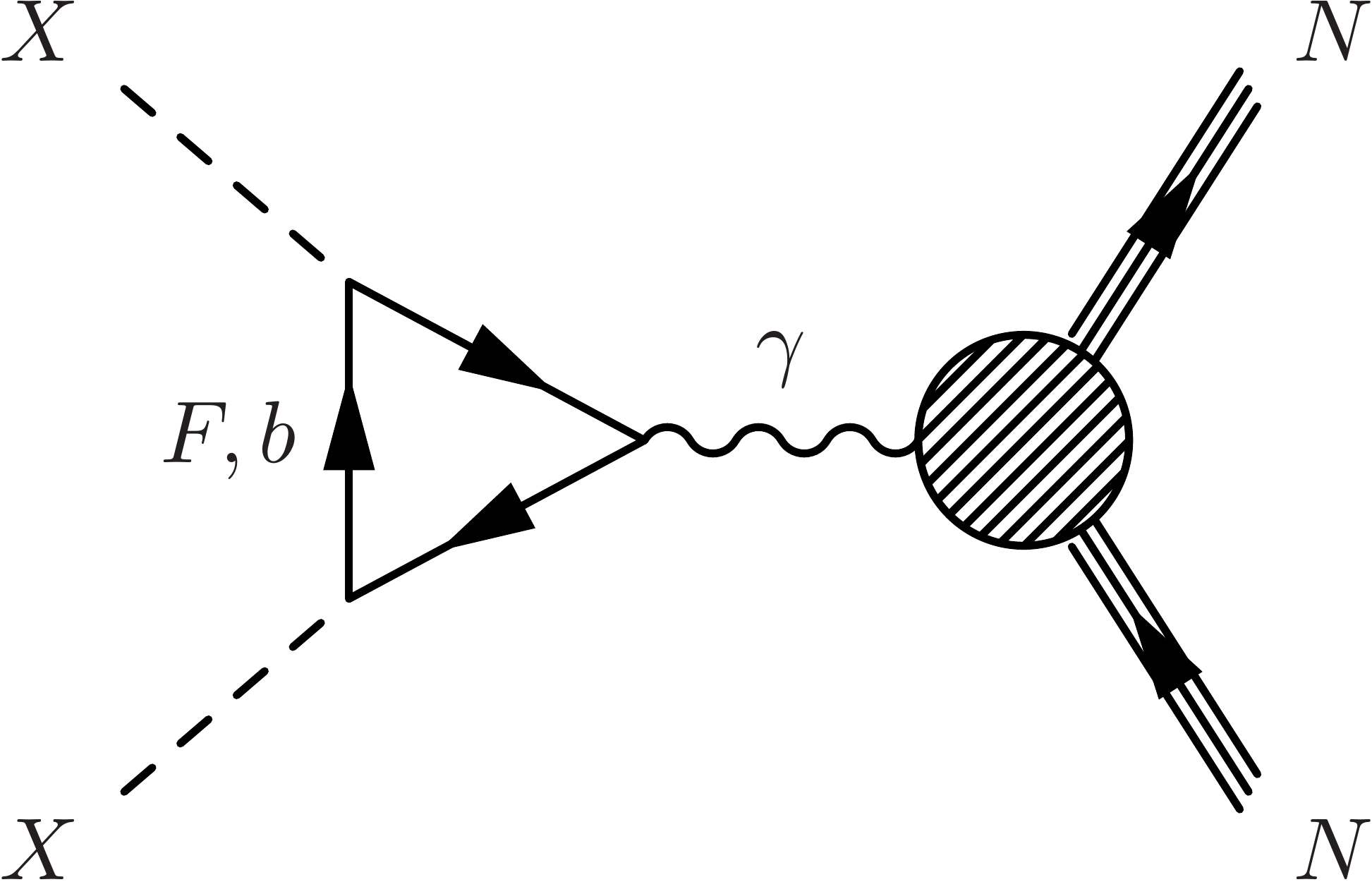,width=0.4\textwidth}}
\caption{The dominant process in the cross section for the direct detection of $X$.
 }
\label{diagram2}
\end{figure}

With the assumption, $\lambda_b \gg |\lambda_d|, \, |\lambda_s|$, 
the tree-level contribution of 
the $s$-channel exchange of the bottom partner to 
the direct detection cross section is sufficiently small 
and the significant contribution arises from the one-loop diagrams, as shown in Fig. \ref{diagram2}. 
Note that there is a logarithmic enhancement, $\log(m_b^2/m_F^2)$, in the $m_F \gg m_b$ region\cite{FlavoredDM1,FlavoredDM2,FlavoredDM3}.
DM can also scatter off gluons in the nucleon via $F$ and $b$ box diagrams. 
The contribution of the gluon scattering is sub-dominant and less than 10\% of the photon exchanging process in our parameter space.
Our analysis includes both the one-loop contributions. 
We use the central limits given by the LUX \cite{LUX2015,LUX2016} and Panda experiments \cite{Panda} to find excluded region.
The red regions in Fig. \ref{fig1} are excluded by the direct detection.

Let us comment on the bound from the DM experiments concerned with the indirect detections. 
$X$ and $X^\dagger$ existing in our universe 
annihilate into $b$ and $\overline b$. The constraint on the cross section has been reported by the Fermi-LAT
collaboration \cite{Ackermann:2015zua}. Assuming that $s$-wave contribution is dominant in DM annihilations,
the lower bound on the DM mass reaches about 200 GeV  \cite{Ackermann:2015zua}.
In our model, the annihilation is dominated by the $p$-wave
contribution, which is much smaller at the present temperature than the one at the freeze-out temperature.
Then, the bound from the indirect detection is not strict in our model, as far as
$\lambda_H$ is enough small and the process shown in Fig. \ref{diagram1} dominates over the Higgs exchanging process.
Note that one very strict exclusion limit on the annihilation cross section has been recently reported in Refs. \cite{Cuoco:2016eej,Cui:2016ppb}, based on the latest result of the AMS-02 experiment \cite{AMS-02}.
If we consider the other setups, as discussed in Sec. \ref{sec;fermionicDM}, 
the upper bound on the annihilation cross section of DM to $b \, \overline b$ has already excluded the parameter space where the relic density can be explained. 
On the other hand, our setup can achieve the correct relic density without any conflicts with the direct/indirect detections.

\subsubsection{Triviality bound}
\label{sec;Triviality bound}

In general, large Yukawa couplings bring cutoff scale to models because Yukawa couplings are asymptotic non-free.
This cutoff scale is known as triviality bound. 
We calculate the triviality bound in this model
because the large $\lambda_b$ is required to reproduce the correct relic abundance of DM in our model as we can see from Fig.~\ref{fig1}.
The beta functions for $\lambda_b$ and the QCD coupling are given as follows:
\begin{align} 
\beta_{\lambda_b} & \simeq
\frac{\lambda_b}{(4\pi)^2}
\Big(-8 g_{3}^{2}  + 4 {\lambda_b^2}   \Big)
,
\\
\beta_{g_3} & \simeq  
\frac{1}{(4 \pi)^2}
\left(- \frac{19}{3} g_{3}^{3} \right),  
\end{align}
where $g_3$ is the QCD gauge coupling which is large and cannot be ignored.
We estimate the triviality bound by solving the renormalization group equation with the beta functions at the one-loop level. 
We fill the regions where the triviality bound is below 1000 TeV with green color in Fig.~\ref{fig1}.

\subsection{Flavor Physics}
\label{sec;flavor}
Finally, we investigate flavor physics based on our results in Secs. \ref{sec;LHC} and \ref{sec;DM}.
In our model, Flavor Changing Neutral Currents (FCNCs) are induced by the Yukawa couplings between quarks and the dark matter at the one-loop level.
Since the chiralities of the quarks are right-handed in the Yukawa coupling, we find that the new physics contributions to the flavor violating processes are strongly suppressed.

\begin{table}
\begin{center}
  \begin{tabular}{|c|c||c|c|} \hline
 $\alpha_s(M_Z)$ & $0.1193(16)$ \cite{PDG}&   $\lambda$& 0.22537(61)  \cite{PDG}   \\ 
 $G_F$  & 1.1663787(6)$\times 10^{-5}$ GeV$^{-2}$  \cite{PDG}   &  $A$& $0.814^{+0.023}_{-0.024}$  \cite{PDG}  \\ 
   $m_{b}$&4.18$\pm 0.03$ GeV  \cite{PDG} &  $\overline{\rho}$& 0.117(21)  \cite{PDG} \\ 
  $m_t$& 160$^{+5}_{-4}$ GeV  \cite{PDG} &   $\overline{\eta}$& 0.353(13)  \cite{PDG}  \\ 
      $m_{c}$&1.275$\pm 0.025$ GeV  \cite{PDG}  &  &  \\  \hline
    $m_K$ & 497.611(13) MeV  \cite{PDG}  & $m_{B_s}$ & 5.3663(6) GeV \cite{PDG}  \\ 
     $F_K$ & 156.1(11) MeV \cite{Lattice1} & $m_{B}$ & 5.2795(3) GeV \cite{PDG} \\ 
     $\Hat B_K$ & 0.764(10) \cite{Lattice1}  &  $F_{B_s}$ & 227.7 $\pm$ 6.2 MeV \cite{Lattice1} \\ 
  $(\Delta M_K)_{\rm exp}$  & 3.484(6)$\times 10^{-12}$ MeV  \cite{PDG} & $F_{B}$ &190.6 $\pm$ 4.6 MeV \cite{Lattice1}   \\ 
    $|\epsilon_K|$ & $(2.228(11)) \times 10^{-3}$  \cite{PDG}  &  $\Hat B_{B_s}$ & 1.33(6) \cite{Lattice1} \\ 
     $\eta_1$ & 1.87(76) \cite{Brod:2011ty} & $\Hat B_{B}$ & 1.26(11)  \cite{Lattice1}  \\ 
        $\eta_2$& 0.5765(65) \cite{Buras:1990fn}  & $\eta_B$ & 0.55 \cite{Buras:1990fn}  \\ 
  $\eta_3$ & 0.496(47) \cite{Brod:2010mj}  &  & \\  \hline
  \end{tabular}
 \caption{The input parameters relevant to our analyses. The CKM matrix, $V_{CKM}$, is written in terms of $\lambda$, $A$, $\overline{\rho}$ and $\overline{\eta}$ \cite{PDG}.}
  \label{table;input}
  \end{center}
\end{table}

\subsubsection{$\Delta F=2$ processes}

\begin{figure}[!h]
\centering
{\epsfig{figure=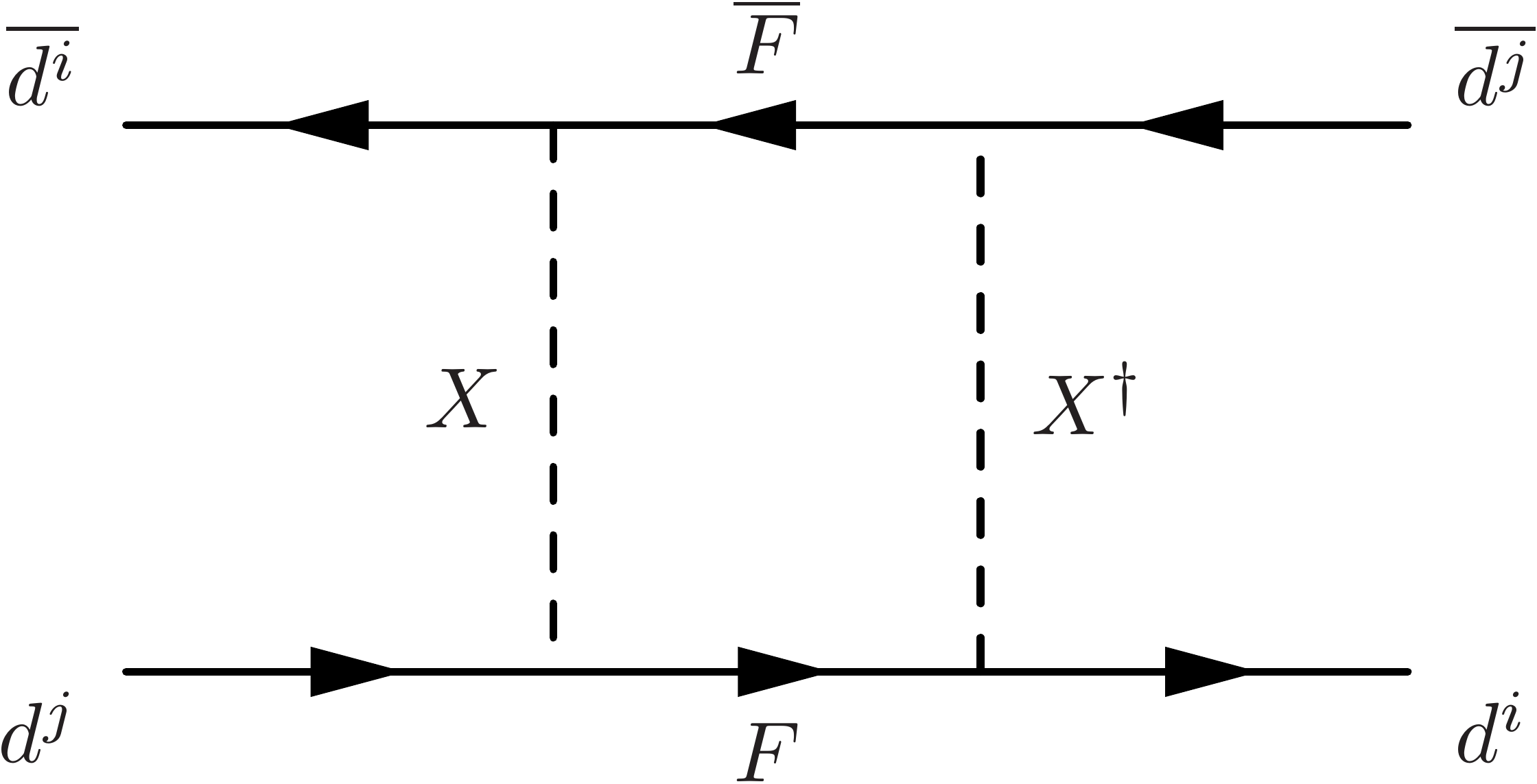,width=0.4\textwidth}}
\caption{The box diagram to contribute to the $\Delta F=2$ processes.
 }
\label{diagram3}
\end{figure}

In the massless limit of the SM quarks, the box diagrams involving $X$ and $F$, shown in Fig. \ref{diagram3},
 induce the operators relevant to the $\Delta F=2$ processes:
\beq
{\cal H}^{\Delta F=2}_{eff}= (\widetilde C_1)_{ij} (\overline{d^i_R} \gamma^\mu d^j_R)  (\overline{d^i_R} \gamma_\mu d^j_R) +h.c..
\eeq
The Wilson coefficients at the one-loop level are given by,
\beq
(\widetilde C_1)_{ij}= \frac{( \lambda_{ j}  \lambda^*_i  )^2}{64 \pi^2} \frac{1}{(m^{ 2}_{F}-m^2_X)^2}
\left \{ \frac{m^{ 2}_{F}+m^2_X}{2} + \frac{m^2_X m^{ 2}_{F}}{m^{ 2}_{F}-m^2_X} \ln\left ( \frac{m^2_X}{m^{2}_{F}} \right ) \right \}.
\eeq

The $K_0$-$\overline{K_0}$, $B_d$-$\overline{B_d}$, and $B_s$-$\overline{B_s}$ mixing are well investigated theoretically and experimentally. Since $\lambda_b$ is ${\cal O}(1)$ as shown in Fig. \ref{fig1}, $B_d$-$\overline{B_d}$ and $B_s$-$\overline{B_s}$ mixing become important even if $|\lambda_d|$ and $|\lambda_s|$ are small compared to $\lambda_b$.
Besides, the physical observables associated with $K_0$-$\overline{K_0}$, in general, constrain new physics contributions,
although their SM predictions still have large uncertainties (See e.g. \cite{Charles:2013aka}). 

Here, we investigate our predictions in the following observables:
\beq
\label{deltaF}
\Delta M_{B_d},~\Delta M_{B_s},~S_{\psi K},~S_{\psi \phi},~\epsilon_K.
\eeq
We do not include $\Delta M_K$, because of the large theoretical ambiguity. 
Among our parameters summarized in Eq. (\ref{parameters}), we expect that 
$m_F$, $m_X$ and $\lambda_b$ are determined by the observables in the DM physics and the LHC experiments.
Then, the other parameters, $Re(\lambda_s),~Im(\lambda_s),~ Re(\lambda_d)$ and $Im(\lambda_d)$,
are fixed by the observables in Eq. (\ref{deltaF}). 
The number of the parameters is smaller than the one of the observables,
so that we can obtain an explicit prediction for the physical quantities measured by the flavor experiments.


In the $B_d$-$\overline{B_d}$ and $B_s$-$\overline{B_s}$ mixing,
the representative observables relevant to the mixing 
are mass differences denoted by $\Delta M_{B_d}$ and $\Delta M_{B_s}$.
They are influenced by $(\widetilde C_1)_{bd}$ and $(\widetilde C_1)_{bs}$ as follows:
\beq
\Delta M_{B_q}= 2 \left |  M^{B_q}_{12} \right |^2 =2 \left | (M^{B_q}_{12})^*_{\rm SM} + \frac{1}{3} (\widetilde C_1)_{bq} m_{B_q} F^2_{B_q} \Hat B_{B_q} \right |^2 ~(q=d, \, s),
\eeq
where $ (M^{B_q}_{12})_{\rm SM}$ is given by the top-loop contribution:
\beq
 (M^{B_q}_{12})^*_{\rm SM}=\frac{G^2_F}{12 \pi^2} F^2_{B_q} \Hat{B}_{B_q} m_{B_q} M^2_W  \{ (V_{CKM})^*_{tb} (V_{CKM})_{tq} \}^2 \eta_BS_0(x_t) .
\eeq
$S_0(x)$ is defined in Appendix \ref{sec;appendix1}. 

The time-dependent CP violations, $S_{\psi K}$ and $S_{\psi \phi}$, are evaluated as follows including the new physics contributions:
\beq
S_{\psi K}=- \sin \varphi_{B_d}, ~ S_{\psi \phi}=\sin \varphi_{B_s},
\eeq
where $ \varphi_{B_q}$ is the phase of $M^{B_q}_{12}$: $M^{B_q}_{12}=|M^{B_q}_{12}|e^{i \varphi_{B_q}}$.
The input parameters are summarized in Table \ref{table;input}, and the central values are used in our analyses.

\begin{figure}[!t]
\centering
{\epsfig{figure=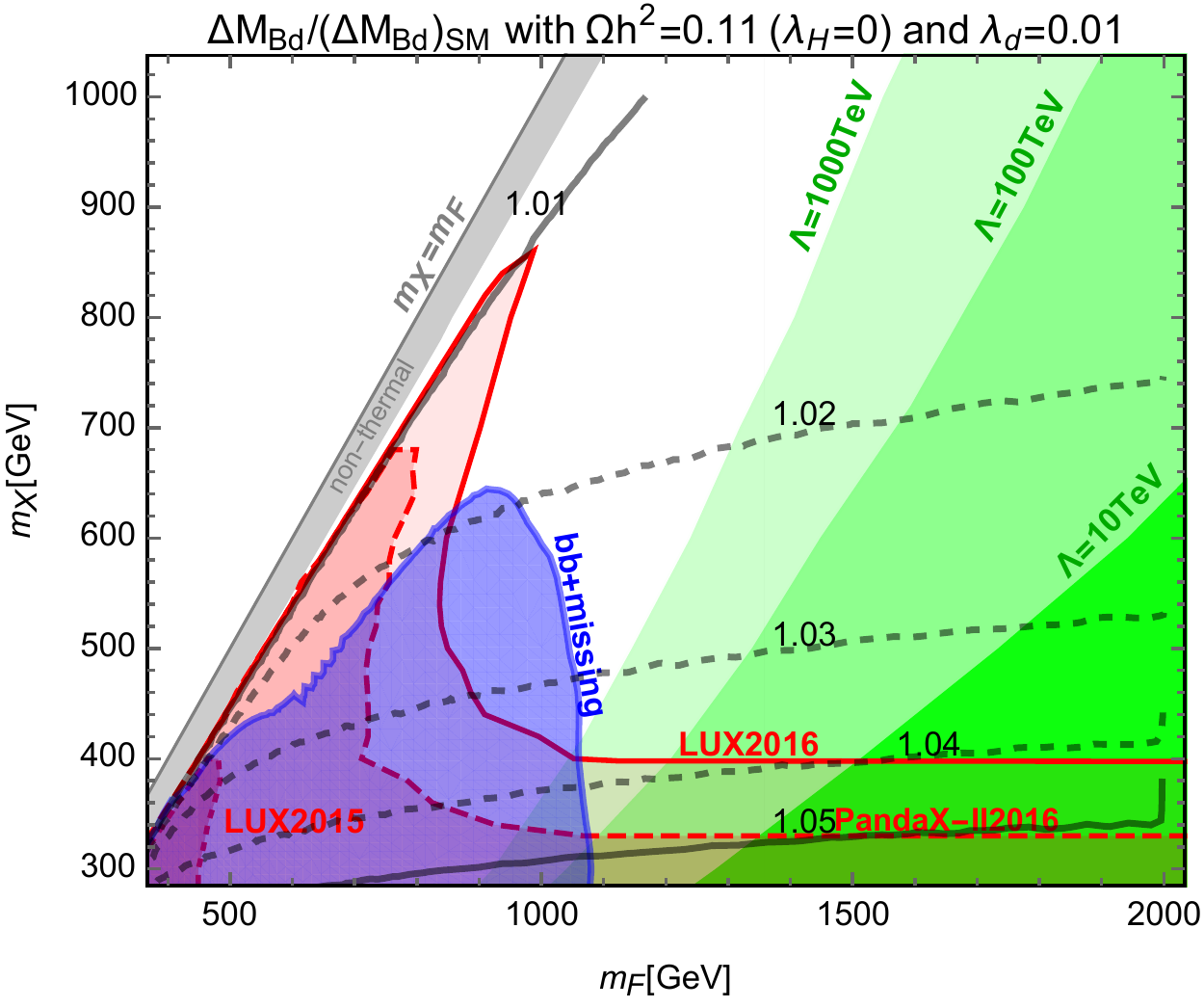,width=0.5\textwidth}}{\epsfig{figure=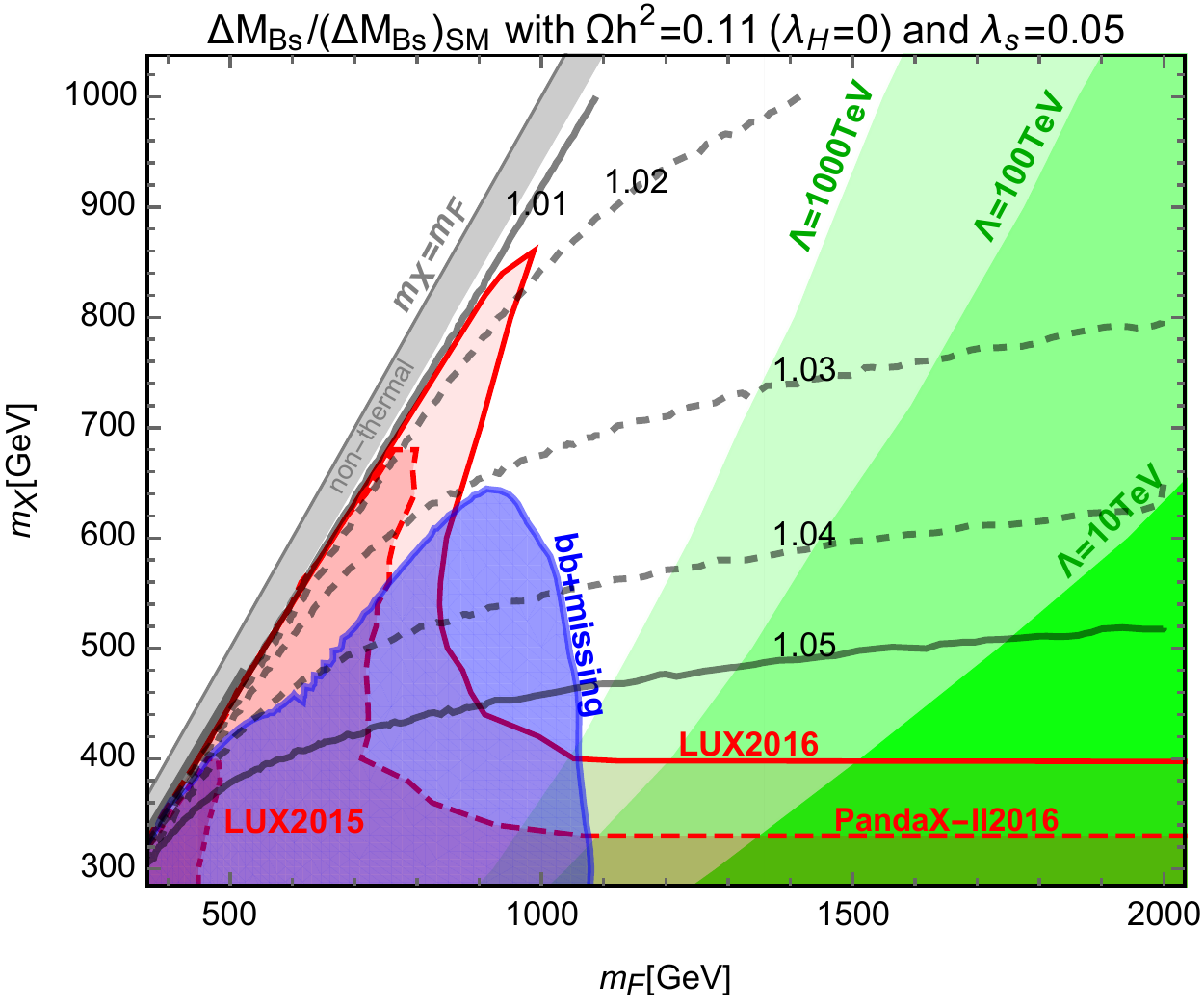,width=0.5\textwidth}}
\vspace{-0.5cm}
\caption{Predictions for the $\Delta F=2$ processes. $\Omega h^2=0.1198 \pm 0.0015$ is satisfied. The blue region is excluded by the
$bb+E_T^{\rm miss}$ at the LHC.
 }
\label{fig2-1}
\end{figure}

In Fig. \ref{fig2-1}, we can see the deviations of $\Delta M_{B_d}$ and $\Delta M_{B_s}$ from the SM predictions,
fixing $\lambda_{d, \,s}$ at $\lambda_d=0.01$ (left panel) and $\lambda_s=0.05$ (right panel).
The solid lines predict 1\% and 5\% deviations respectively, compared to the SM predictions.
The dotted lines correspond to the 2\%, 3\%, and 4\% deviations from bottom to top in each panel.
As we see in those figures, the deviations are enough small to evade the bounds on the $\Delta F=2$ processes,
as far as $|\lambda_d| \leq 0.01$ and $|\lambda_s| \leq 0.05$ are satisfied.
Note that there are still large uncertainties of the SM predictions for $\Delta M_{M_q}$, and 
the CKMfitter collaboration suggests that 10 \% deviations are still allowed according to 
the global analyses \cite{Charles:2013aka,Charles:2015gya}.
If $\lambda_{d}$ ($\lambda_{s}$) is set to 0.02 (0.1), the deviations become about four times bigger than the values on Fig. \ref{fig2-1}.
Then, we could conclude that the upper bounds on $|\lambda_{d}|$ and $|\lambda_{s}|$ are $\cal{O}$(0.01) and $\cal{O}$(0.05) respectively, in the region that the Landau poles do not appear below 1000 TeV.


\begin{figure}[!t]
\centering
{\epsfig{figure=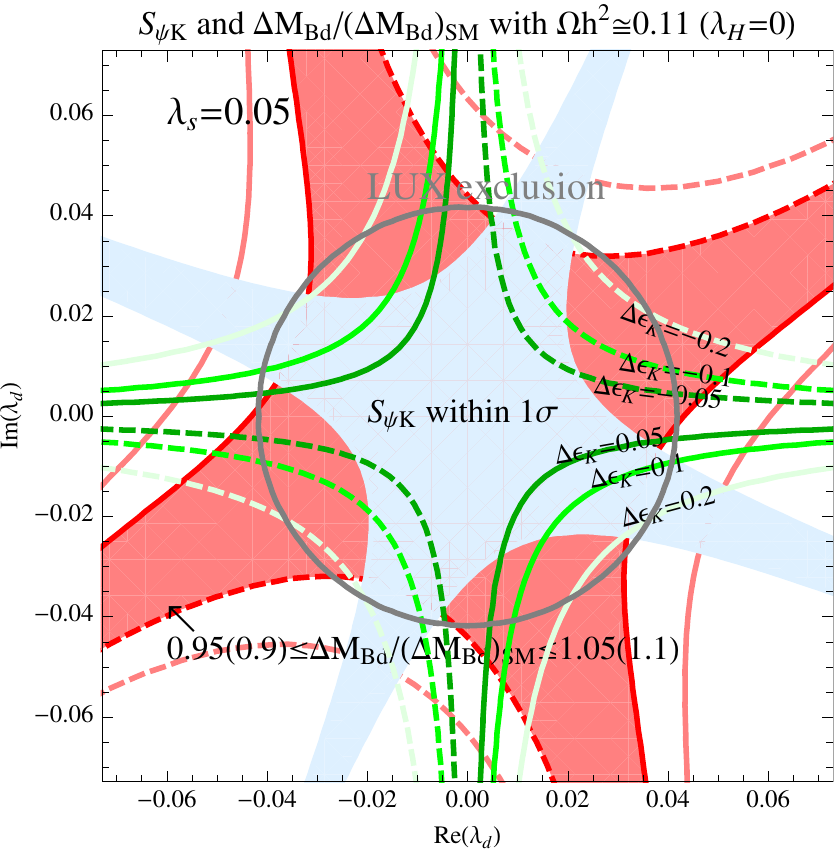,width=0.5\textwidth}}{\epsfig{figure=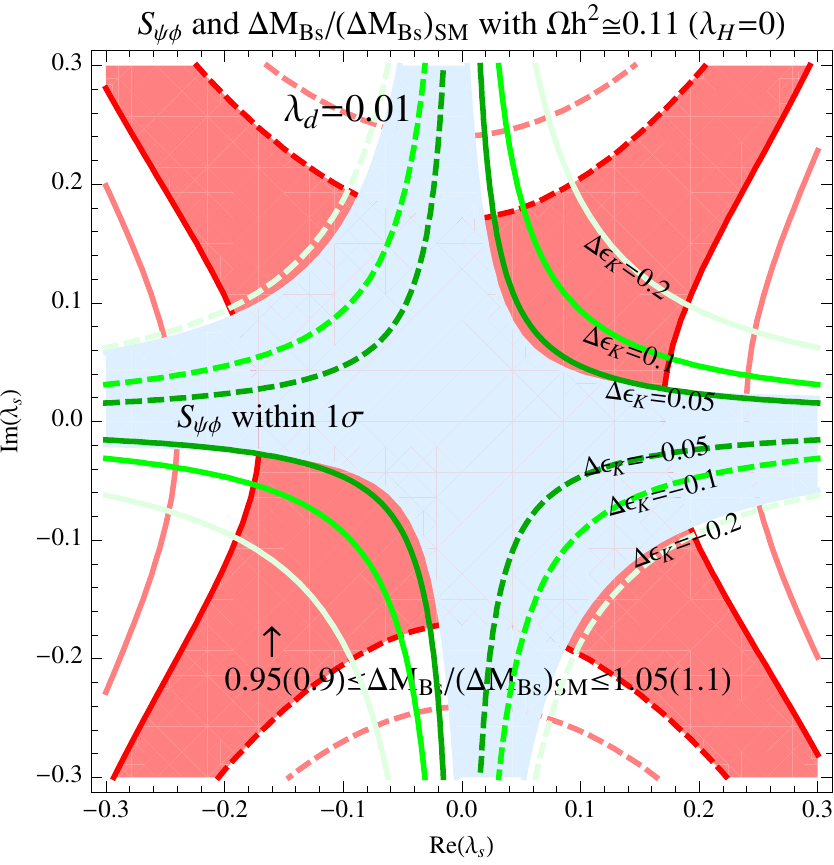,width=0.5\textwidth}}
\vspace{-0.5cm}
\caption{Predictions for the $\Delta F=2$ processes. The other parameters are fixed by  the relic abundance within 1$\sigma$: $(m_X,\, m_F, \, \lambda_b)=(900\, {\rm GeV}, \, 964.4 \, {\rm GeV}, \,0.66)$. The gray circle depicts the exclusion of the LUX experiment \cite{LUX2016}. The outside of the circle exceeds the upper bound on the cross section of 
the DM direct detection.
 }
\label{fig2}
\end{figure}

\begin{figure}[!t]
\centering
{\epsfig{figure=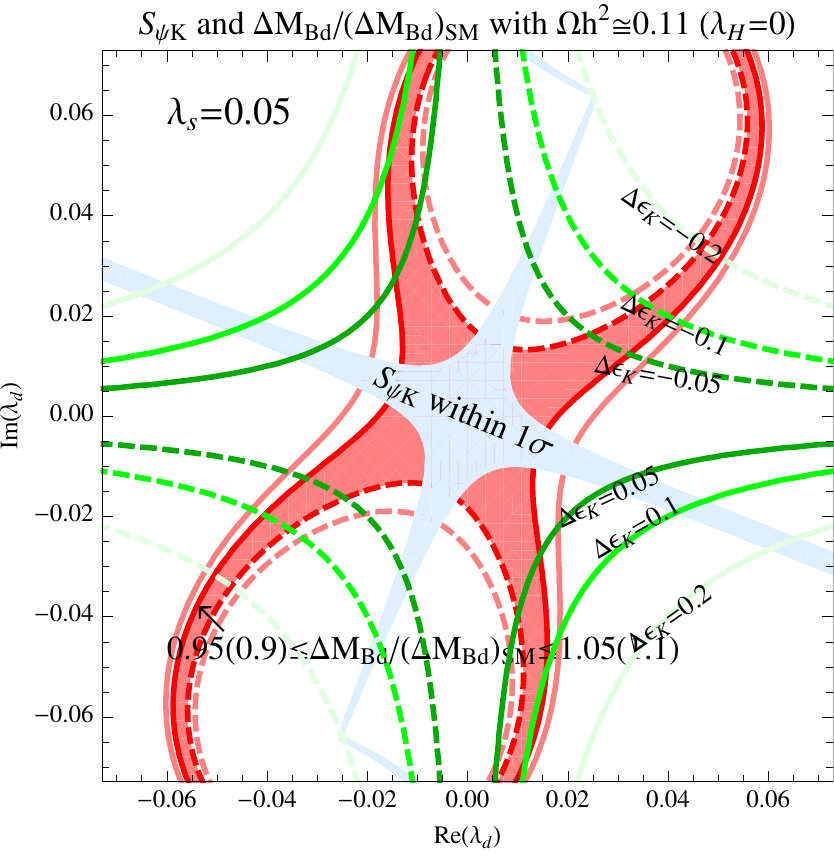,width=0.5\textwidth}}{\epsfig{figure=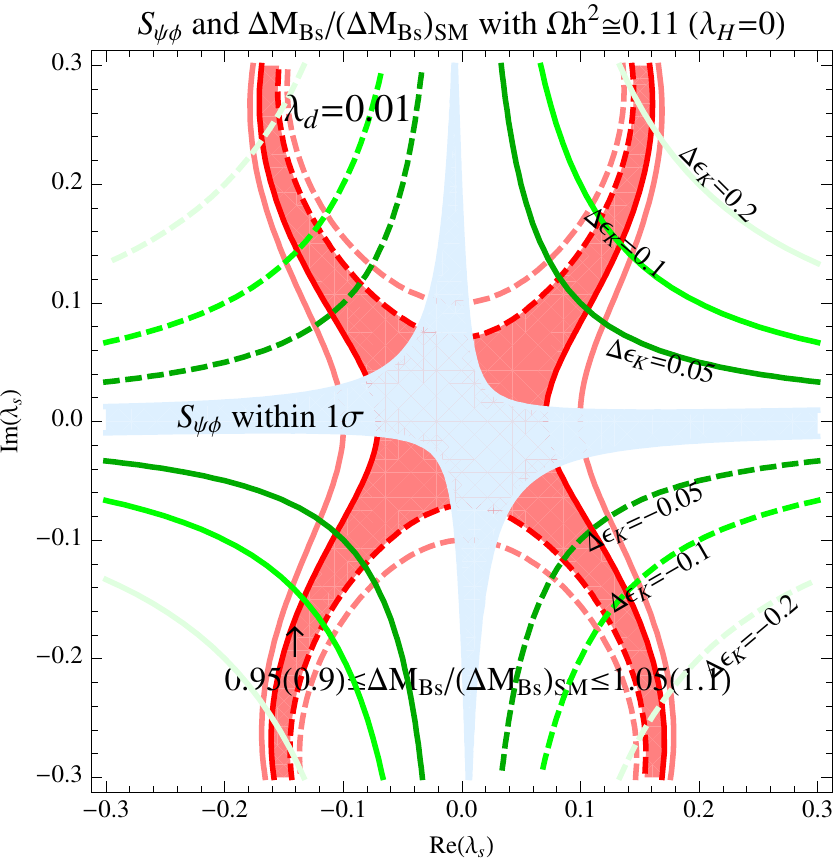,width=0.5\textwidth}}
\vspace{-0.5cm}
\caption{Predictions for the $\Delta F=2$ processes. The other parameters are fixed by the relic abundance within 1$\sigma$: $(m_X,\, m_F, \, \lambda_b)=(900\, {\rm GeV}, \, 1795.7 \, {\rm GeV}, \,2.32)$. The exclusion lines proposed by the LUX experiment \cite{LUX2016} are out of these parameter regions. 
 }
\label{fig3}
\end{figure}

In Fig. \ref{fig2} and Fig. $\ref{fig3}$, we can see the bounds on the $\Delta F=2$ processes, more clearly.
We fix $m_X$, and $m_F$ and $\lambda_b$, according to the requirement of the correct relic density within 
$1 \, \sigma$. We choose the reference points: $(m_X,\, m_F, \, \lambda_b)=(900\, {\rm GeV}, \, 964.4 \, {\rm GeV}, \,0.66)$, $(900\, {\rm GeV}, \, 1795.7 \, {\rm GeV}, \,2.32)$ in Fig. \ref{fig2} and Fig. \ref{fig3} respectively. 
On the blue bands, the deviations of $S_{\psi K}$ and $S_{\psi \phi}$ are within $1 \sigma$ : $S_{\psi K}=0.691 \pm 0.017$ and $S_{\psi \phi}= 0.015 \pm 0.035$ \cite{PDG}.
On the red (dashed) lines, the deviations of $\Delta M_{M_q}$ are 5 \% (-5 \%). The pink (dashed) lines 
predict 10 \% (-10 \%) deviations, respectively. In the red regions, the magnitudes of the deviations 
are less than 5 \%.
We see that the region where the magnitudes of the deviations are less than 5 \% for the $\Delta F=2$ processes corresponds to
$|\lambda_d|  \lesssim 0.04$ (left panel) and $|\lambda_s|  \lesssim 0.2$ (right panel) in Fig. \ref{fig2}.
$m_F$ and $\lambda_b$ in Fig. \ref{fig3} are bigger than in Fig. \ref{fig2}.
Such a large $m_F$ tends to suppress deviations of the observables, but large $\lambda_b$ is required to
achieve the correct relic density. 
Thus, the allowed region becomes narrow for the large $m_F$. 

We can find the correlation between the flavor physics and the DM direct detection, especially in
the left panel of Fig. \ref{fig2}. If $|\lambda_d|$ is sizable, the tree-level diagram, $X^\dagger \, d \to X^\dagger d $,  induces significant deviations. 
The gray circle is the exclusion line of the LUX experiment \cite{LUX2016}, which was discussed in Sec. \ref{sec;DMdirect}. Then, we see that the allowed region roughly corresponds to $|\lambda_d|\lesssim 0.04$.
In the right panel of Fig. \ref{fig2} and Fig. \ref{fig3}, the exclusion line is out of the figure.
Note that the upper bound from the direct detection is roughly $|\lambda_d|\lesssim 0.15$, in the right panel of Fig. \ref{fig3}.


The observables of the $K_0$-$\overline{K_0}$ mixing can be estimated. Their SM predictions are described by 
$(M^K_{12})_{\rm SM}$,
\beq
(M^K_{12})_{\rm SM}^*= \frac{G^2_F}{12 \pi^2} F^2_K \Hat{B}_K m_K M^2_W \left \{ V^2_c \eta_1S_0( x_c) +  V^2_t \eta_2 S_0(x_t) + 2  V_c V_t \eta_3 S(x_c, x_t)  \right \},
\eeq
where $x_i \equiv m^2_i/M^2_W$ and $V_i \equiv (V_{CKM})^*_{is} (V_{CKM})_{id}$ are defined, respectively. $\eta_{1,2,3}$ correspond to the NLO and NNLO QCD corrections. Each function is defined in Appendix \ref{sec;appendix1}
and the used values are summarized in Table \ref{table;input}.
The physical observables on the $K_0$-$\overline{K_0}$ mixing are denoted by $\epsilon_K$ and $\Delta M_K$,
which are described as 
\beq
\epsilon_K= \frac{\kappa_\epsilon e^{i \varphi_\epsilon} }{\sqrt{2} (\Delta M_K)_{\rm exp}} \, Im(M^K_{12}), ~ \Delta M_K =2  Re(M^K_{12}).
\eeq
$\kappa_\epsilon$ and $\varphi_\epsilon$ are given by the observations:
$\kappa_\epsilon=0.94 \pm 0.02$ and $\varphi_\epsilon=0.2417 \times \pi$.
$M^K_{12}$ includes the new physics contribution and is decomposed as follows in our model:
\beq
(M^K_{12})^*=\left ( M^K_{12} \right )^*_{\rm SM} + ( \widetilde{ C}_1)_{sd} \times \frac{1}{3} m_K F^2_K \Hat B_K .
\eeq
The running correction is included at the one-loop level in our analysis.

The predictions for the deviations of $\epsilon_K$ are depicted as green lines in Figs. \ref{fig2} and $\ref{fig3}$.
Once the deviations of the $B_d$-$\overline{B_d}$ and $B_s$-$\overline{B_s}$ mixing
are discovered, we can principally predict the deviation of $\epsilon_K$.
The (dashed) dark, normal and light green lines depict the (-)5\%, (-)10\%, and (-)20\% deviations of $\epsilon_K$ respectively, compared to the SM prediction. 
On each panel, $\lambda_s=0.05$ (left) and $\lambda_d=0.01$ (right) are assumed.
In Fig. \ref{fig2}, they correspond to  $(\Delta M_{B_s}/(\Delta M_{B_s})_{\rm SM}, \, S_{\psi \phi})=(1.004, \, 0.037)$ (left) and $(\Delta M_{B_d}/(\Delta M_{B_d})_{\rm SM}, \, S_{\psi K})=(1.003, \, 0.687)$ (right). 
In Fig. \ref{fig3}, the fixed $\lambda_s$ and $\lambda_d$ correspond to  $(\Delta M_{B_s}/(\Delta M_{B_s})_{\rm SM}, \, S_{\psi \phi})=(1.025, \, 0.037)$ (left) and $(\Delta M_{B_d}/(\Delta M_{B_d})_{\rm SM}, \, S_{\psi K})=(1.015, \, 0.678)$ (right) respectively. Besides, the current limit on the direct detection of DM constraints the 
deviation of $\epsilon_K$, depending on the mass region. As we see in the left panel of Fig. \ref{fig2},
$|\Delta \epsilon_K|$ cannot exceed about $0.2$, in this compressed mass region.

\subsubsection{$b \, \to s \, \gamma$ and the other observables}

The $b \to s$ transitions may be good processes to test our model.
The contributions of the new Yukawa couplings are, however, too small to find the deviations in flavor experiments.
The structure of the chirality suppresses the photon- and $Z$-penguin diagrams.
The chirality-flipped operators are suppressed by the quark masses on the external lines.
One of the most important processes to test new physics is $b \to s \, \gamma$.
The relevant operators are 
\beq
{\cal H}^{b \to s \gamma}_{eff}= C_7 (\overline{s_L} \sigma^{\mu \nu} b_R) F_{\mu \nu}+C'_7 (\overline{s_R} \sigma^{\mu \nu} b_L) F_{\mu \nu}.
\eeq  
In our setup, the new contribution to $C'_7$ is larger than the one to $C_7$,
because of the mass difference between $m_b$ and $m_s$.
The allowed new physics contribution is well summarized in Ref. \cite{Descotes-Genon:2013wba}:
$|C'_{7}| \lesssim 0.02$. Fixing $\lambda_s$ at $\lambda_s=0.1$, we estimate $|C'_{7}|$ as
$|C'_{7}| \lesssim 0.004$, as far as the deviation of $\Delta M_{B_s}$ is less than 10 \%.
Then, we conclude that it is difficult to test our model, using the $b \to s \, \gamma$ process.

The other processes associated with the $b \to s$ transition 
may constrain our model. It is interesting that some excesses in the observables of  $B \to K^* \, l  \, l$
have been reported,
but the $Z$-penguin diagram that contributes to the $b \to s$ transition is vanishing in our model.
Therefore, we conclude that the $b \to s$ transition is not so relevant to our model.

\section{Comparison with the Dirac DM case }
\label{sec;fermionicDM}

\begin{figure}[!t]
\centering
{\epsfig{figure=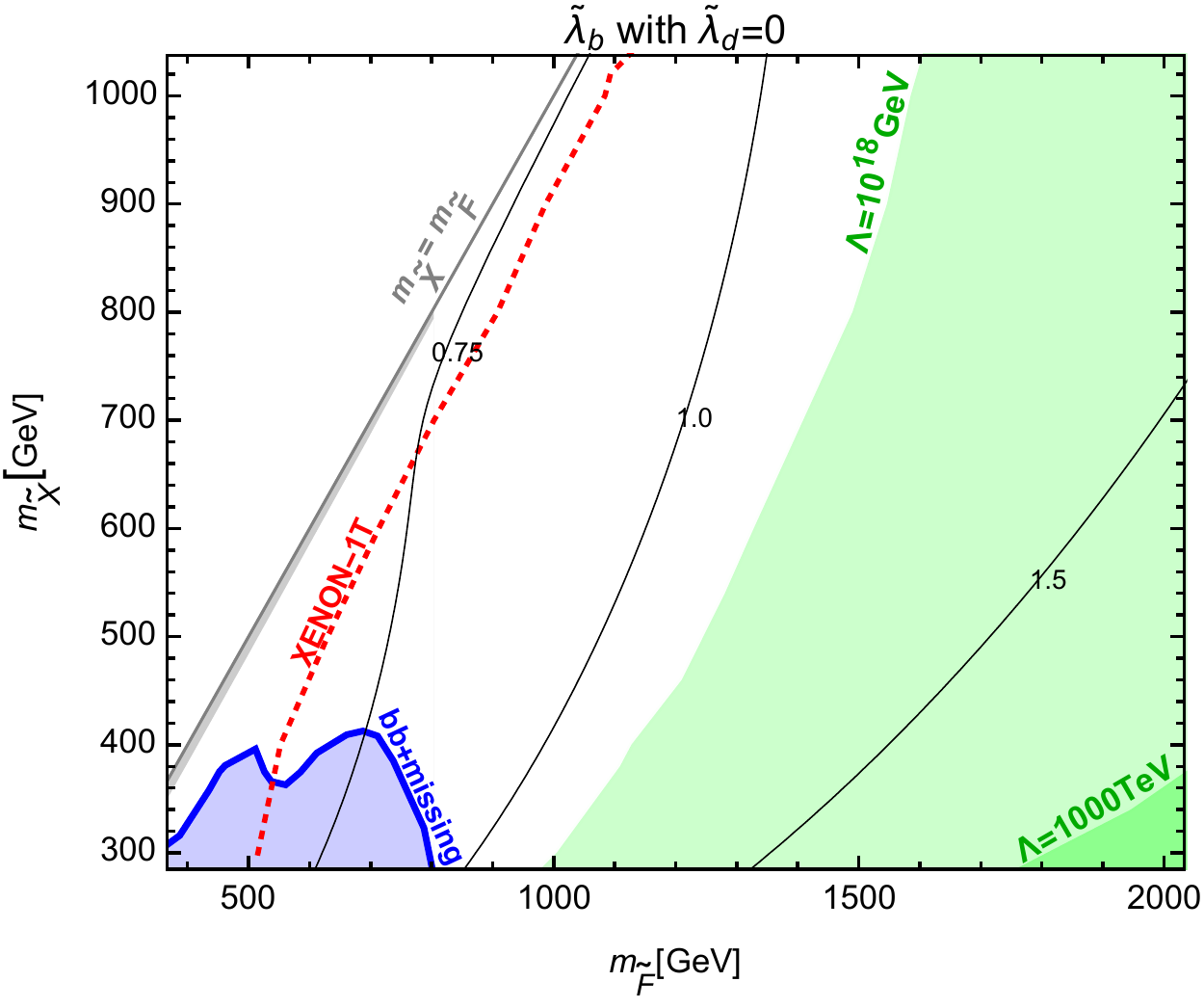,width=0.7\textwidth}}
\vspace{-0.5cm}
\caption{The required values of $\widetilde \lambda_b$ for the correct relic abundance in the Dirac-fermion DM model with an extra colored scalar particle. The blue region is excluded by the
$bb+E_T^{\rm miss}$ at the LHC.
 }
\label{fig4}
\end{figure}
It is important to see differences among the predictions of DM models with extra colored particles, as well.
As discussed above, one of the stringent constraints is from the direct detection of the complex scalar DM.
This is because the $s$-wave contribution of the annihilation cross section of DM through the $t$-channel process
is suppressed by the fermion mass, and then ${\cal O}$(1) Yukawa coupling, $\lambda_b$, is required to achieve the correct relic density.
If we introduce a gauge singlet Dirac-fermion DM $(\widetilde X)$ as a candidate of DM 
with an extra colored scalar field $(\widetilde F)$ instead of $X$ and $F$, we can expect that the $s$-wave contribution is enhanced by the DM mass.
The Yukawa couplings relevant to the annihilation are given by
\beq
\widetilde \lambda_{ i} \widetilde F^\dagger  \overline{\widetilde X_L} d^i_{R} + h.c.
\eeq

Figure \ref{fig4} shows the required value of $\widetilde \lambda_b$ for the correct relic abundance of DM within $1 \sigma$.
As mentioned above, the $s$-wave contribution is efficient to reduce the abundance and
then $\widetilde \lambda_b$ is relatively small in the parameter region.
This leads the small cross section for the direct detection of $\widetilde X$,
so that even XENON-1T could cover only the compressed region.
The blue region is excluded by the $bb+E_T^{\rm miss}$ at the LHC, 
where the production cross section of the extra scalars
at the LO exceeds the experimental upper bounds
using the same data as the case of extra fermions.
Note that the parameter region of Fig. \ref{fig4} seems to face the stringent constraint from the latest AMS-02 result \cite{Cuoco:2016eej,Cui:2016ppb}.

We also estimate the triviality bound.
The beta functions for $\tilde{\lambda}_b$ and the QCD coupling are given as follows.
\begin{align} 
\beta_{\widetilde \lambda_b} & \simeq
\frac{1}{(4\pi)^2}
\Big(-4 g_{3}^{2}  + 3 {\widetilde \lambda_b^2}  \Big) \widetilde \lambda_b
,
\\
\beta_{g_3} & \simeq  
\frac{1}{(4 \pi)^2}
\left(- \frac{41}{6} g_{3}^{3} \right). 
\end{align}
We fill the region where the triviality bound is below $10^{18}$~GeV and 1000~TeV with green color in Fig.~\ref{fig4}.
The bound is weaker than in the scalar DM case 
because the required Yukawa coupling to reproduce the correct relic abundance is smaller than in the scalar DM case.



\section{Summary }
\label{sec;conclusion}
The existence of DM is one of mysteries which could be solved by the
extension of the SM.
There are a lot of possibilities of the extensions, and we seriously have to 
examine what kind of extended SMs can explain the DM abundance
in our universe. 
The Weakly Interacting Massive Particle (WIMP) scenario is one of the popular setups for DM, 
and WIMP DM models could be classified in terms of interactions between DM and quarks/leptons.
If DM interacts with the SM particles via the electroweak gauge couplings,
the gauge interactions would be dominant and effective to achieve the relic abundance of DM.
If DM is a SM gauge singlet, new interactions would be required as far as the Higgs exchanging is not so efficient.
In this paper, we focus on a possibility that 
dark matter mainly annihilates through the t-channel exchange of the extra quark into pairs of SM quarks.
Interestingly, the new interaction is flavor-dependent, so that this simple DM model can be tested 
by flavor physics as well as DM physics and the LHC experiments. 
If DM signals are confirmed in the direct/indirect detections of DM or/and the LHC experiments, 
we have to find out a promising DM model among many candidates,  using independent physical observables. 
In our model, we can expect some correlations and explicit predictions in observables of flavor physics, 
so that our simple model can be tested by the accurate measurements of the flavor violating processes. 
In fact, the region, where the DM relic abundance is explained, is very close to the exclusion limit of the DM direct detections in the scalar DM case, 
so that our DM may be discovered near future. 
Besides, the Belle II experiment will start in 2018 and the measurements of the new physics contributions to flavor physics, 
e.g. the $\Delta F=2$ processes,
are expected to be drastically improved \cite{Charles:2013aka}. Then, our models can be tested via 
the observations of the $\Delta F=2$ processes with the great accuracies. Our setup is very simple and
predicts distinguishing deviations from the SM predictions in flavor physics. 
Moreover, the thermal relic abundance of the DM suggests the large Yukawa coupling of the DM with bottom quark,
so that the deviations in the $\Delta F=2$ processes are sizable. 
Note that a discrepancy of the observables in the $\Delta F=2$ processes has been proposed in Ref. \cite{Blanke:2016bhf}. It will be important to discuss the consistency with the observations, considering the discrepancy \cite{progress}.

In order to compare with another DM model, we also present our results in a Dirac-fermion DM case.
There are still a lot of possible setups which are not studied here: real scalar DM case, $top$ $partner$ model and so on.
It is very important to clearly understand the differences among them and to prepare for the discovery of DM. We have to find out how to test and distinguish DM models. The study for the comparison will be pursued in future. Note that our model presented here is one of the realist setups to evade the stringent bound from the indirect detection of DM \cite{Cuoco:2016eej,Cui:2016ppb}.


\section*{Acknowledgments}
We are grateful to C. Boehm, A. J. Buras and A. Ibarra for comments and suggestions.
This work was supported by JSPS KAKENHI Grant Number 16K17715 [TA].
The work of J. K. was supported by Grant-in-Aid for
Research Fellow of Japan Society for the Promotion of
Science No. 16J04215.

\appendix

\section{Functions}
\label{sec;appendix1}
The functions which appear in $K_0$-$\overline{K}_0$ mixing are given by
\begin{eqnarray}
S_0(x)&=&  \frac{4x -11x^2+x^3}{4(1-x)^2} - \frac{3x^3 \log x}{2(1-x)^3}, \\
S(x,y)&=&\frac{-3xy}{4(y-1)(x-1)} - \frac{xy(4-8y+y^2) \log y}{4(y-1)^2(x-y)}  \nonumber \\
&&+\frac{xy(4-8x+x^2) \log x}{4(x-1)^2(x-y)}.
\end{eqnarray}




\end{document}